\documentclass[prl,nofootinbib,twocolumn]{revtex4}

\usepackage{color}
\usepackage{graphicx}
\usepackage{amsmath}
\usepackage{hyperref}

\begin{document}

\title{Information Gain in Cosmology: \\ From the Discovery of Expansion to Future Surveys}

\author{Marco Raveri$^{1,2,3}$, Matteo Martinelli$^{4,5}$, Gong-Bo Zhao$^{6,7}$ and Yuting Wang$^{6,7}$}
\affiliation{
\smallskip
$^{1}$ SISSA - International School for Advanced Studies, Via Bonomea 265, 34136, Trieste, Italy \\
\smallskip
$^{2}$ INFN, Sezione di Trieste, Via Valerio 2, I-34127 Trieste, Italy \\
\smallskip
$^{3}$ INAF-Osservatorio Astronomico di Trieste, Via G.B. Tiepolo 11, I-34131 Trieste, Italy \\
\smallskip
$^{4}$ Institute Lorentz, Leiden University, PO Box 9506, Leiden 2300 RA, The Netherlands \\
\smallskip
$^{5}$ Institut f\"ur Theoretische Physik, Ruprecht-Karls-Universit\"at Heidelberg, Philosophenweg 16, 69120 Heidelberg, Germany.  \\
\smallskip
$^{6}$ National Astronomy Observatories, Chinese Academy of Science, Beijing, 100012, P.R.China \\
\smallskip
$^{7}$ Institute of Cosmology \& Gravitation, University of Portsmouth, Dennis Sciama Building, Portsmouth, PO1 3FX, UK
}

\begin{abstract}
Facing the advent of the next generation cosmological surveys we present a method to forecast knowledge gain on cosmological models. We propose this as a well defined and general tool to quantify the performance of different experiments in relation to different theoretical models. In particular, the assessment of experimental performance will benefit enormously from the fact that this method is invariant under re-parametrization of the model. We apply this to future surveys and compare expected knowledge advancements to the most relevant experiments performed over the history of modern cosmology. When considering the standard cosmological model, we show that it will rapidly reach knowledge saturation in the near future and forthcoming improvements will not match the past ones. On the contrary, we find that new observations have the potential for unprecedented knowledge jumps when extensions of the standard scenario are considered.
\end{abstract}

\maketitle

Over the last century, modern cosmology saw enormous theoretical and technological advancements that brought it to the era of  ``precision cosmology''. 
After its introduction, the standard $\Lambda$-Cold Dark Matter cosmological model ($\Lambda$CDM) saw remarkable experimental confirmation and nowadays all the parameters defining this model are all measured with high precision. \\
Still, many aspects of this model are lacking a deep physical interpretation. In particular, regardless of the increasing accuracy of cosmological measurements, the nature of Dark Matter and Dark Energy is still unclear.
Future surveys are designed and are being developed to push the experimental precision with which we map the universe even further, with the aim of shedding light on these aspects. \\
With this data abundance awaiting us in the near future it is becoming increasingly relevant to quantify knowledge advancements. 
Future cosmological experiments are requiring a longer and longer time to be developed and we need to plan ahead the scientific targets of such missions. \\
With the present {\it letter} we show how to quantify and forecast the knowledge gain from cosmological probes and we propose this as a well defined and general Figure of Merit to quantify experiment performances.
We apply this statistical technique to some of the most relevant experiments performed over all the history of modern cosmology and we forecast future knowledge advancements.
We show the improvement of the information brought to cosmology by observations, from Hubble's measurements to the most up to date available data and, following the same procedure, we quantify the contribution of upcoming planned experiments. We perform this investigation both assuming the $\Lambda$CDM model, with minimal extensions, and allowing for more freedom in the Dark Energy sector. \\
We show that the $\Lambda$CDM model is rapidly approaching knowledge saturation in the near future with an information gain that does not match the advancements of the past century.
On the contrary, when considering extensions of the $\Lambda$CDM model, we still have in the future significant knowledge jumps that will improve the knowledge on these models when compared to the present day state and the past. \\

{\it Information Gain as a figure of merit.}

In the quest for a way of quantifying and forecasting knowledge gain we shall resort to information theory. In particular we shall use the Kullback-Leibler divergence, also known as information gain or relative entropy that quantifies the proximity of two probability distributions.
Consider two probability density functions (PDF), $P_1$ and $P_2$ of a $d$ dimensional random variable $\theta$. The Kullback-Leibler (KL) divergence is defined by:
\begin{align} \label{Eq:KLDivergence}
D\left( P_2 || P_ 1 \right) \equiv \int P_2(\theta) \log_{2}\left(\frac{P_2(\theta)}{P_1(\theta)} \right)  \, d\theta \hspace{0.5cm} \mbox{[ bits ]}
\end{align}
and represents the information difference in going from $P_1$ to $P_2$, in bits~\cite{Kullback:1951va}.
The KL divergence finds application in several branches of science and was used in a cosmological context in~\cite{Kunz:2006mc, Paykari:2012ne, Amara:2013swa, Seehars:2014ora, Verde:2014qea, Grandis:2015qaa, Seehars:2015qza}. \\
Here we apply Eq.~(\ref{Eq:KLDivergence}) to the posterior of two different experiments by setting $P_{1,2} = P(\theta|\mathcal{M})\mathcal{L}(\theta, D_{1,2}, \mathcal{M})$, where $P(\theta|\mathcal{M})$ denotes our prior on the $\theta$ cosmological parameters of model $\mathcal{M}$ and $\mathcal{L}(\theta, D_{1,2}, \mathcal{M})$ stands for the likelihood of the two data sets respectively, within model $\mathcal{M}$. \\
We focus on forecasting this quantity and to do so we assume that the posterior of the two considered experiments is Gaussian both in the cosmological parameters and the data. This is clearly an optimistic assumption whose goodness depends on the constraining power of the considered probe. 
Since we focus on optimistic forecasting of cosmological knowledge improvement, the Gaussian assumption fits in with this ideal set up.
We shall call $F_{1,2}$ the Fisher matrices of the two data sets and identify those with the inverse parameters covariance of the likelihood. In addition we shall consider a prior distribution characterized by a Fisher matrix $F_p$. In this set-up we assume that the mean parameters of the posterior are the same for all the considered distributions; in doing this we focus on the information improvement resulting from the shrinking of the posterior distribution, neglecting the contribution due to a difference in the mean parameters. \\
Under these assumptions it can be shown that the KL divergence~(\ref{Eq:KLDivergence}) can be written as:
\begin{align} \label{Eq:KLDivergenceForecast}
D\left( P_2 || P_ 1 \right)  = \frac{1}{2\ln 2} \bigg[ & -\ln \frac{\det \left( F_1+F_p\right) }{\det \left( F_2+F_p \right)} \nonumber \\
&-d +{\rm Tr}\left[ \left( F_2+F_p\right)^{-1} \left( F_1+F_p \right) \right]\bigg]. \nonumber \\
\end{align}
Using this quantity to measure the performances of a cosmological experiment has several advantages over other statistical indicators, such as the Dark Energy Task Force (DETF) Figure of Merit (FoM)~\cite{Albrecht:2006um}. We shall briefly comment here on the main ones.
The first advantage of the KL divergence is that it is invariant under reparametrizations, as can be seen from its definition Eq.~(\ref{Eq:KLDivergence}) or its Gaussian approximation, Eq.~(\ref{Eq:KLDivergenceForecast}).
That is, within a given model, the quantity of information that we posses does not depend on our, arbitrary, choice of the parameters definition.
In order to understand this invariance, let us assume the original Fisher matrices are mapped into new ones using a similarity transformation, 
\begin{align}
\tilde{F_1} = (J^{-1} {F_1}^{-1} J)^{-1} \hspace{0.2cm} ; \hspace{0.2cm}  \tilde{F_2} = (J^{-1} {F_2}^{-1} J)^{-1}.
\end{align}
Substituting the above equations into Eq.~(\ref{Eq:KLDivergenceForecast}), using the identity $\ln det A = Tr \ln A$ and the fact that the trace is invariant under similarity transformations, we can easily see that Eq.~(\ref{Eq:KLDivergenceForecast}) is invariant under a reparametrisation transformation. 
Another advantage is that Eq.~(\ref{Eq:KLDivergenceForecast}) is well defined in the case of degenerate parameters. Thanks to these properties it can be used to compare consistently the performances of a cosmological probe for different models on the same scale.
Finally it incorporates a balancing factor, dependent on the dimension $d$ of the model parameter space, that penalizes models with many parameters, if no knowledge is gained on those. \\
In the remaining of this {\it Letter} we show three applications of the KL divergence. The first one consists in quantifying the information gained by a single experiment with respect to the prior distribution, and we shall call it the {\it Prior Information Gain}. 
The second one consists in quantifying the information gain, with respect to the prior, of multiple experiments, adding all the independent ones progressively and replacing the surveys whenever a more recent observation of the same probe is available. We shall call this the {\it Cumulative Information Gain}.
The third application that we show consists in evaluating the information gained only on a particular aspect of a model, in our case the parameters defining an extension of the $\Lambda$CDM model. To do so we marginalize over the base $\Lambda$CDM parameters and we evaluate the cumulative information gain from the Fisher matrix of the model parameters. We shall refer to this as the {\it Model Information Gain}.

{\it Data sets, models and tools.} 

\begin{figure*}[!t]
\begin{center}
\includegraphics[width=1\textwidth]{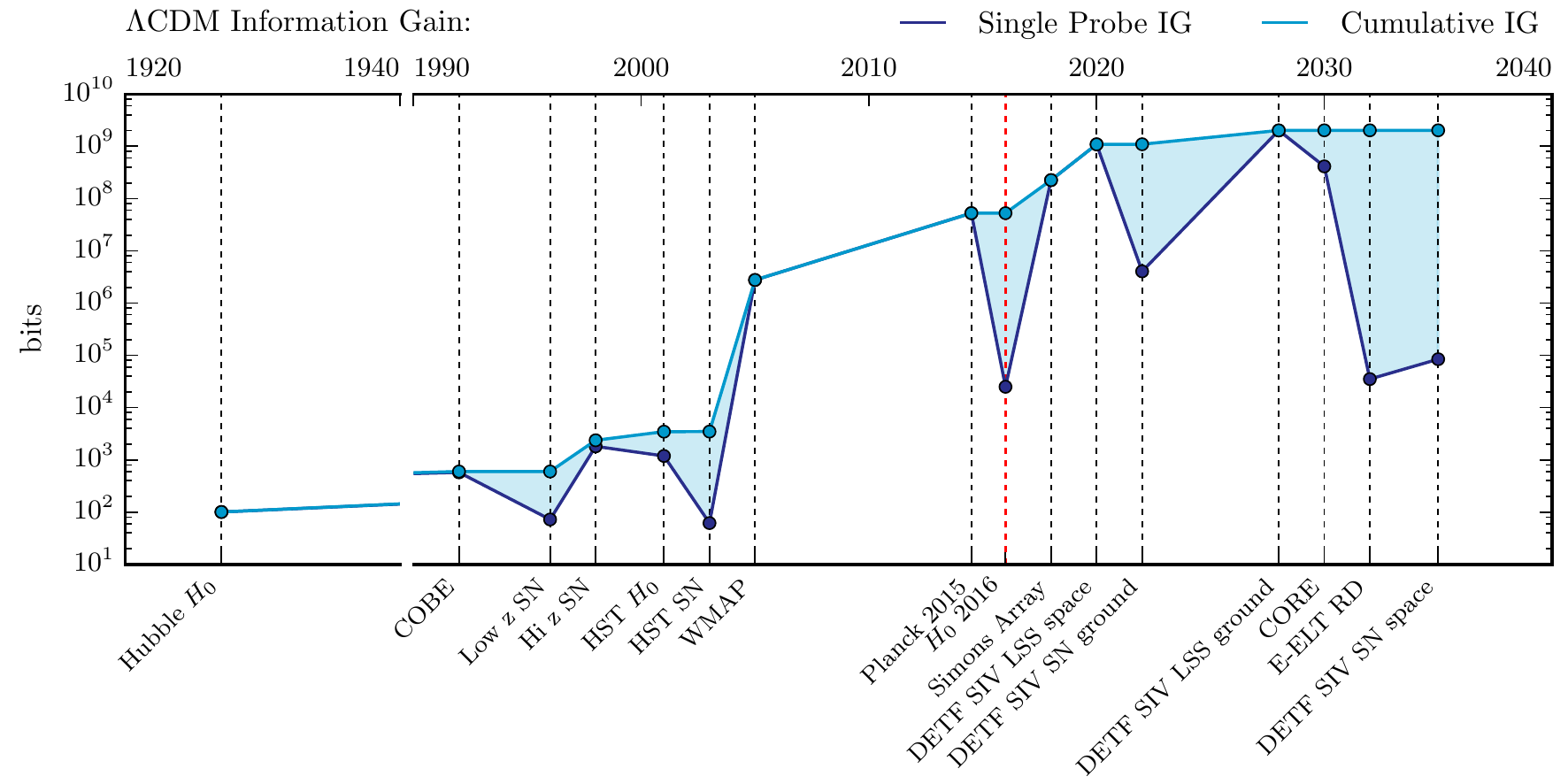}
\caption{{\it Information Gain on the $\Lambda$CDM model.} The light curve displays the cumulative ammont of information obtained by summing all the independent experiments up to a given time. The dark curve measures the information content of each of the single probes. The red dashed curve indicates the present time.}\label{Fig:LCDMIG}
\end{center}
\end{figure*}

In this {\it Letter} we shall compute the information gained, on different cosmological models, by several past and future experiments. We shall consider the following past probes: the determination of the Hubble constant, as measured by Edwin Hubble~\cite{Hubble:1929ig}, as determined by the Hubble Space Telescope (HST) Key Project~\cite{Freedman:2000cf} and the HST in 2016~\cite{Riess:2016jrr}; measurements of the Cosmic Microwave Background (CMB) anisotropies by the Cosmic Background Explorer (COBE)~\cite{Smoot:1992td,Mather:1993ij}, Wilkinson Microwave Anisotropy Probe (WMAP)~\cite{Bennett:2012zja} and {\it Planck}~\cite{Adam:2015rua} satellites; measurements of the luminosity distances of low redshift~\cite{Betoule:2014frx} and high redshift supernovae~\cite{Riess:1998cb, Perlmutter:1998np} and from the HST Supernova Survey~\cite{Riess:2006fw, Suzuki:2011hu}. \\
In addition to these experiments we shall consider the following future surveys: measurements of CMB anisotropies by the Simons Array~\cite{Errard:2015cxa} and by the Cosmic ORigins Explorer (CORE) satellite~\cite{011arXiv1102.2181T}; the DETF specifications for Stage IV LSS and supernovae survey both ground and space based\footnote{the space based LSS survey is modified to account for a larger sky coverage $f_{\rm sky}=0.5$}~\cite{Albrecht:2006um}; measurements of redshift drift from the European-Extremely Large Telescope~\cite{Liske:2008ph}. As we place ourselves in a ideal forecasting framework, we always consider the most optimistic specifications available in the above references.
For LSS probes we consider both weak lensing shear and galaxy clustering measurements, accounting for their cross correlations and for the cross correlation with CMB probes. \\
The time placement of future surveys is subject to errors and, while the scheme that we choose is probably unrealistic, we argue that the physical picture does not strongly depend on that.

We consider the information that is gained and will be obtained on several models.
The first that we consider is the baseline $\Lambda$CDM model, as specified by its six parameters $( \Omega_b h^2, \Omega_c h^2, h, {\rm{ln}}(10^{10} A_s), n_s, \tau)$, respectively the fractional density of baryons, the fractional density of CDM, the reduced Hubble constant $h=H_0/100$, the amplitude of the primordial scalar perturbations spectrum and its spectral index, the reionization optical depth.
Then we consider some extensions of the standard picture: the $\Lambda$CDM model with the addition of massive neutrinos~\cite{Lesgourgues:2006nd}; the $\Lambda$CDM model with the addition of a primordial gravitational waves component, parametrized by the tensor to scalar ratio $r$; the Chevallier-Polarski-Linder (CPL) parametrization of the Dark Energy equation of state~\cite{Chevallier:2000qy,Linder:2002et}; the gravitational growth index $\gamma$ parametrization for a modified growth of cosmic structures~\cite{Linder:2007hg}; Low-energy Ho\v rava gravity, that was first proposed in~\cite{Horava:2008ih} and included in EFTCAMB in~\cite{Frusciante:2015maa}. We consider the model in Ho\v rava gravity that evades solar system constraints~\cite{Audren:2014hza,Frusciante:2015maa}.

All the results presented in this {\it Letter} are obtained with the CosmicFish forecasting code~\cite{CosmicFishNotes}. This allows to obtain Fisher matrix forecast for all the above considered cosmological experiments and models. To obtain cosmological predictions for the $\Lambda$CDM, $\Lambda$CDM + $m_{\nu}$, $\Lambda$CDM + $r$, CPL models the CosmicFish code uses the CAMB sources code~\cite{Lewis:1999bs, Challinor:2011bk}, for the $\gamma$ parametrization of the modified growth of cosmic structures it uses MGCAMB~\cite{Zhao:2008bn, Hojjati:2011ix}, for EFT based parametrizations of DE/MG, as well as single DE/MG models like Ho\v rava gravity, it uses the EFTCAMB code~\cite{Hu:2013twa, Raveri:2014cka}.

The CosmicFish code is publicly available at~\url{http://cosmicfish.github.io}. A CosmicFish package containing all the relevant code to automatically produce all the results presented in this {\it Letter} is publicly available as well in the same website.

{\it Information Gain in $\Lambda$CDM.} 

\begin{figure*}[!t]
\begin{center}
\includegraphics[width=1\textwidth]{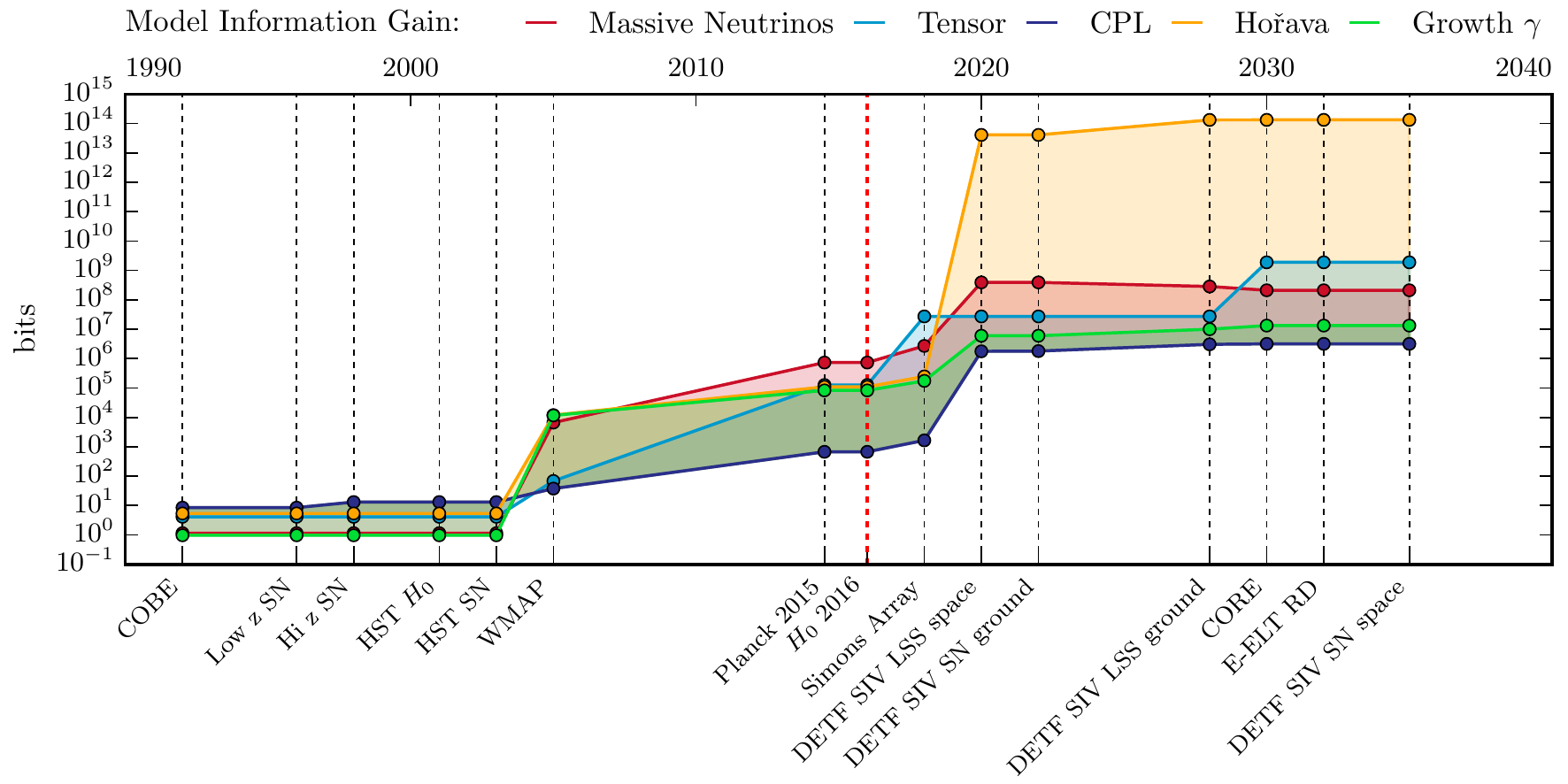}
\caption{{\it Model cumulative Information Gain}. Different lines and colors show the cumulative information gained on the parameters defining extensions of the $\Lambda$CDM model, as explained in the legend. The red dashed curve indicates the present time.}\label{Fig:DEMGIG}
\end{center}
\end{figure*}

We start by analyzing in depth the information gained in the standard $\Lambda$CDM model. Since this is our fiducial cosmological model it is particularly relevant to evaluate future knowledge advancements in the light of past ones.
Figure~\ref{Fig:LCDMIG} shows the information progression, quantified with the KL divergence Eq.~(\ref{Eq:KLDivergenceForecast}), along the history of modern cosmology as spanned by different experiments. \\
As we can see in this figure the cumulative information gain, obtained by combining all the independent available experiments at a given time, shows several interesting features. 
At first we see that we have noticeable information jumps corresponding to measurements of the COBE satellite followed by Hi redshift supernovae, and by the HST $H_0$ measurements. 
A radical gain in information comes from the WMAP satellite that, in this picture, corresponds to the probe that gained most of the knowledge on the $\Lambda$CDM model. The {\it Planck} satellite further improved this knowledge by one order of magnitude and the Simons Array is expected to substantially contribute as well.
After the Simons Array the next jump in knowledge correspond to space based Stage IV LSS surveys that add to the picture the information coming from the clustering of cosmic structures. Noticeably after this probe the information about the $\Lambda$CDM model saturates and there is no other substantial improvement. \\
The most remarkable result for the $\Lambda$CDM model is that the information gained in the last century, that amounts to roughly 5.5 orders of magnitude, is not matched by future improvements that will raise our knowledge about the universe by just 1.5 orders of magnitude. \\
Figure~\ref{Fig:LCDMIG} also shows the information gained by all probes considered alone, i.e. what we call Single Probe IG. This allows us to compute what is the best probe for the $\Lambda$CDM model and, not surprisingly, it is a LSS Stage IV survey from ground, followed by the LSS Stage IV survey from space and CMB probes. \\
We can see that, as time passes, the information content of all probes increases, in particular when we consider the same kind of probe. $H_0$ and SN measurements, in particular, improve by orders  of magnitude but are somewhat penalized when compared to the constraining power of CMB and LSS probes.

{\it Information Gain in extended models.}

The physical picture previously presented drastically changes when considering extended and alternative models to the $\Lambda$CDM one. 
Figure~\ref{Fig:DEMGIG} shows the information gain, quantified with the KL divergence Eq.~(\ref{Eq:KLDivergenceForecast}), as spanned by different, time ordered, experiments.
This figure in particular focuses on the information gain of different cosmological probes when considering only the parameters defining the extensions of the $\Lambda$CDM model, having marginalized over the base parameters, what we called {\it Model Information Gain}.
As we can see WMAP observations provide a improvement in knowledge on all the considered model, specially on MG models and on massive neutrinos. Similarly the {\it Planck} CMB survey increases information on all $\Lambda$CDM extensions considered. This time, however, measurements of $r$ together with the neutrino mass receive the highest boost in information, thanks to the measurement of CMB polarization. \\
Moving to future experiments, we can see that the Simons Array will only slightly improve the knowledge about DE/MG models while providing a substantial jump on primordial GWs. \\
Models of DE/MG, instead, experience their next substantial knowledge improvement from space based LSS surveys.	
In particular, within the considered models, the information gain on Ho\v rava gravity is particularly significant with a jump of approximately $8$ orders of magnitude.
This huge advancement in knowledge is driven by galaxy clustering and weak lensing. All the LSS windows and their cross correlations are modified and strong constraints arise from this tomographic information. More specifically the bounds on the theory parameters go from $\sim 10^{-4}$ as in~\cite{Frusciante:2015maa} to $\sim 10^{-7}$.
Since these models mainly alter the growth of cosmic structures, after this probe, the information on them saturates and remains constant.
In a similar way, massive neutrinos receive their last significant increase in knowledge from LSS ground based observations. 
On the other hand, the investigation of primordial GWs, will receive a substantial amount of information from the CORE satellite, with a jump of roughly two orders of magnitude on top of Simons Array advancements. \\
Allow us to stress that Figure~\ref{Fig:DEMGIG} can be used to compare the performances of one experiment on different models on the same comparison scale, which is instead not possible with other FoM definitions.
This can be used to understand and prioritize the scientific goals of cosmological experiments improving their aim.

In conclusion, in this {\it Letter}, we have used the KL divergence to quantify knowledge advancements in cosmology. In particular we worked out its application to forecasting and proposed to use it as a well defined, flexible and powerful FoM for cosmological experiments.
Using the KL divergence to measure the performances of an experiment has, in fact, substantial advantages over other FoM that have been proposed in literature. In particular it is invariant under reparametrizations and allows us to compare the knowledge gain on different models on the same comparison scale.
In addition, several applications for this statistical tool are possible, ranging from identifying the most suited experiment for a given cosmological effect to the optimization of the design of future cosmological surveys.
We used the KL divergence to quantify the information gained on different cosmological models by many past and future experiments. 
This shows that, when considering the $\Lambda$CDM model, the knowledge gained in the last century is not matched by future experiments. In addition, information gain will saturate with the introduction of Stage IV LSS surveys from space and experiments after that will not substantially rise our knowledge of the $\Lambda$CDM model. \\
We have shown that this picture is radically different when considering several extensions of the $\Lambda$CDM model: massive neutrinos; primordial GWs; the CPL parametrization of the DE EoS; the $\gamma$ parametrization of a modified growth of structures; Ho\v rava gravity.
For the aspects characterizing these models we have shown that future surveys will result in key knowledge improvements, and in particular for Ho\v rava models these will have an enormous jump.

These results show that, while in the last century we measured the parameters of the $\Lambda$CDM model to high precision, in the future the gain in information will be connected to the discovery of new physics, outside the description of the standard cosmological model.

The results presented in this {\it Letter} were possible thanks to the forecasting tools that we implemented in the CosmicFish code that is publicly available at~\url{http://cosmicfish.github.io}.

\vskip 10pt

\begin{acknowledgments}
We are grateful to Ana Ach\'ucarro, Carlo Baccigalupi, Erminia Calabrese, Stefano Camera, Luigi Danese, Giulio Fabbian, Noemi Frusciante, Bin Hu, Valeria Pettorino, Levon Pogosian, Giuseppe Puglisi and Alessandra Silvestri for useful and helpful discussions on the subject. We are indebted to Luca Heltai for help with numerical algorithms.
MM is supported by the Foundation for Fundamental Research on Matter (FOM) and the Netherlands Organization for Scientific Research / Ministry of Science and Education (NWO/OCW). MM was also supported by the DFG TransRegio TRR33 grant on The Dark Universe during the preparation of this work.
MR acknowledges partial support by the Italian Space Agency through the ASI contracts Euclid-IC (I/031/10/0) and the INFN-INDARK initiative.
MR acknowledges the joint SISSA/ICTP Master in High Performance Computing for support during the development of this work.
MR thanks the National Astronomical Observatories, Chinese Academy of Science for the hospitality during the initial phases of development of this work.
MR and MM thank the Galileo Galilei Institute for Theoretical Physics for the hospitality and the INFN for partial support during the completion of this work.
GBZ and YW are supported by the Strategic Priority Research Program ``The Emergence of Cosmological Structures'' of the Chinese Academy of Sciences Grant No. XDB09000000, and by University of Portsmouth. YW is supported by the NSFC grant No. 11403034. 
\end{acknowledgments}

\end{document}